\documentstyle[amssymb,12pt,aps,prbbib,eqsecnum,preprint]{revtex}
\input{epsf}
\tightenlines
\begin{document}
\title{Vacuum effects in an asymptotically uniformly accelerated frame with a
constant magnetic field}
\author{V\'{\i}ctor M. Villalba \thanks{%
e-mail:villalba@ivic.ivic.ve}}
\address{Centro de F\'{\i}sica\\
Instituto Venezolano de Investigaciones Cient\'{\i}ficas, IVIC\\
Apdo 21827, Caracas 1020-A, Venezuela}
\author{Juan Mateu \thanks{%
e-mail mateu@ciens.ula.ve}}
\address{Centro de Investigaciones de Astronom\'{\i}a CIDA, \\
Urb. Alto Chama II etapa. Calle 8. Qta. CIDA. Merida, Venezuela}
\maketitle

\begin{abstract}
In the present article we solve the Dirac-Pauli and Klein Gordon equations
in an asymptotically uniformly accelerated frame when a constant magnetic
field is present. We compute, via the Bogoliubov coefficients, the density
of scalar and spin 1/2 particles created. We discuss the r\^ole played by
the magnetic field and the thermal character of the spectrum.
\end{abstract}


\section{Introduction}

The study of quantum effects in nonintertial frames of reference has been
thoroughly discussed in the literature. The pioneer articles of Fulling and
Unruh \cite{Fulling,Unruh} showing the non-equivalence of the quantization
of scalar fields in Rindler and Minkowski coordinates and the thermal
character of the radiation were the origin of a large body of articles
devoted to analyze quantum measurement processes in uniformly accelerated
frames and possible interpretations thereof.

The advantage of considering Rindler coordinates are many. They can be
associated with a uniformly accelerated observer. They also possess a global
timelike Killing vector and the (massive and massless) Klein-Gordon as well
as the Dirac equations are separable in the Rindler coordinates. The Rindler
Coordinates can be extended to cover the whole space time and thermal
effects can be related, via the equivalence principle, with the Hawking
effect \cite{Birrel,Grib}.

The study of quantum effects in non uniformly accelerated frames of
reference presents, at first glance, different technical problems. Among
them we can mention that the system of coordinates associated with
non-Rindler Kinematics do not possess in general a timelike Killing vector,
and therefore a standard interpretation of positive and negative frequency
solutions is absent. The complete separation of variables of\ \ the
Klein-Gordon and the Dirac equations is possible only in a restricted set of
coordinates, and those coordinates allowing separability sometimes present
coordinate singularities; therefore the quantization scheme fails.

Among the non-static coordinate systems where the Klein-Gordon and Dirac
equations separate we have\cite{Kalnins,Costa}

\begin{equation}
\begin{array}{c}
t+x=\frac{2}{\omega }\sinh \omega \left( T+X\right) ,\ t-x=\frac{-1}{\omega }%
e^{-\omega \left( T-X\right) },\ y=y,\ z=z
\end{array}
.  \label{1}
\end{equation}
The line element associated with the coordinate transformation (\ref{1}) is 
\begin{equation}
ds^{2}=\left( e^{-2\omega T}+e^{2\omega X}\right) \left(
dX^{2}-dT^{2}\right) +dy^{2}+dz^{2}.  \label{2}
\end{equation}

The separability of he Klein-Gordon equation in (\ref{1}) \ has been
discussed by Kalnins \cite{Kalnins}, \ one the authors has accomplished a
complete separability of the Dirac\cite{Villalba} equation in (\ref{1}).

The kinematics associated with the coordinates (\ref{1}) have been
exhaustively analyzed by Costa \cite{Costa,Costa1}. The proper time along $%
X=X_{0}$ is 
\[
s=\frac{1}{\omega }e^{-2\omega T}+e^{2\omega X_{0}}+\frac{1}{\omega }%
e^{\omega X_{0}}\sinh ^{-1}e^{\omega (T+X_{0})}. 
\]
Considering $T$ as the evolution parameter, we have that an observer
co-moving to the system (\ref{1}) has a four-velocity given by 
\begin{equation}
{\bf V=}\frac{\left( 
\begin{array}{cccc}
\cosh \omega \left( X+T\right) +\frac{e^{\omega \left( X-T\right) }}{2}, & 
\cosh \omega \left( X+T\right) -\frac{e^{\omega \left( X-T\right) }}{2}, & 0,
& 0
\end{array}
\right) }{\sqrt{e^{-2\omega T}+e^{2\omega X}}}
\end{equation}
and experiences an acceleration whose components are 
\begin{equation}
a_{t,x}=\frac{\omega \sinh \omega \left( X+T\right) \mp \omega \frac{%
e^{\omega \left( X-T\right) }}{2}}{e^{-2\omega T}+e^{2\omega X}}+\frac{\cosh
\omega \left( X+T\right) \pm \frac{e^{\omega \left( X-T\right) }}{2}}{\left(
e^{-2\omega T}+e^{2\omega X}\right) ^{2}}\omega e^{-2\omega T},\quad
a_{y}=0,\quad a_{z}=0.  \label{acel}
\end{equation}
From the absolute value of the acceleration, we readily obtain that $%
a=\left| a^{\mu }a_{\mu }\right| ^{1/2}$ takes the form 
\begin{equation}
a=\omega e^{2\omega X}\left( e^{-2\omega T}+e^{2\omega X}\right) ^{-\frac{3}{%
2}}.  \label{3}
\end{equation}
From (\ref{3}) we obtain that the accelerated frame becomes inertial as $%
T\longrightarrow -\infty $, and on the other hand it evolves toward an
uniformly accelerated frame as $T\longrightarrow +\infty $.

Quantum effects in the non-inertial frame (\ref{1}) have been discussed by
Costa \cite{Costa} and by Percoco and Villalba \cite{Percoco} for Dirac
Particles. The nonexistence of a global timelike Killing vector for the line
element (\ref{2}) precludes making a straightforward identification of the
positive and negative frequency solutions of the scalar and Dirac wave
equations. In order to circumvent this difficulty, we identify positive and
negative frequency modes comparing the asymptotic solutions of the wave
equations with those obtained for the relativistic Hamilton-Jacobi equation

Recently, Bautista \cite{Bautista}, has discussed, in a Rindler accelerated
frame, vacuum effect associated with a spin 1/2 particle with anomalous
magnetic moment in a constant magnetic field directed along the
acceleration. The author obtains a Planckian distribution of created
particles that depends on the magnetic field via the non-minimal anomalous
coupling. The author also discusses deviation of the energy density from a
thermal distribution due to the magnetic field. \ As a preliminary step
towards a deeper understanding of quantum processes in non inertial frames
of reference, \ in the present paper we analyze vacuum effects associated
with scalar and spin $1/2$ \ particles in the coordinates (\ref{1}), when a
constant magnetic field is also present. \ \ 

The article is structured as follows: In Section II, we solve the
relativistic Hamilton-Jacobi equation with a constant magnetic field in the
accelerated frame (\ref{1}). In Sec. III, we solve the Klein-Gordon equation
and compute the rate of scalar particles created. In Sec. IV, we solve the
Dirac equation with anomalous magnetic moment and compute the density of
particles related by the magnetic field in the non inertial frame (\ref{1}).
Finally, we discuss the results obtained in this article in Sec. V.

\section{Solution of the Hamilton-Jacobi equation}

The relativistic Hamilton-Jacobi equation coupled to an electromagnetic
field can be written as\cite{Costa,Landau} 
\begin{equation}
g^{\alpha \beta }\left( \partial _{\alpha }S-eA_{\alpha }\right) \left(
\partial _{\beta }S-eA_{\beta }\right) +m^{2}=0  \label{HJ}
\end{equation}
where here and elsewhere we adopt the units where $c=1$, and $\hbar =1.$ \
The vector potential associated with a constant magnetic field \ $\vec{B}%
=B_{x}\hat{x}$ directed along the acceleration (\ref{acel}) has the form

\begin{equation}
{\bf A}=\left( 0,0,B_{x}z,0\right)  \label{A}
\end{equation}
It is not difficult to verify that (\ref{A}) satisfies the conditions $%
\nabla _{\mu }A^{\mu }=0$ and $F^{\alpha \beta }F_{\alpha \beta }=2B_{x}^{2}$%
.

Substituting the line element (\ref{2}) into (\ref{HJ}) we obtain

\begin{equation}
\begin{array}{c}
\frac{1}{e^{2\omega X}+e^{-2\omega T}}\left[ \left( \partial _{X}S\right)
^{2}-\left( \partial _{T}S\right) ^{2}\right] +\left( \omega _{o}z-\partial
_{y}S\right) ^{2}+\left( \partial _{z}S\right) ^{2}+m^{2}=0
\end{array}
\label{k24}
\end{equation}
where $\omega _{o}=eB_{x}$ . The solution of Eq. (\ref{k24}) has the form

\begin{equation}
\begin{array}{c}
S(X,y,z,T)=-k_{y}y\pm i\int \sqrt{m^{2}-\lambda ^{2}+\left( k_{y}+\omega
_{o}z\right) ^{2}}dz\pm \\ 
\pm \int \sqrt{k-\lambda ^{2}e^{2\omega X}}dX\pm \int \sqrt{k+\lambda
^{2}e^{-2\omega T}}dT,
\end{array}
\label{k25}
\end{equation}
which in the asymptotic limit as $X\longrightarrow \infty $ and $%
T\longrightarrow -\infty $ reduces to

\begin{equation}
S(X,y,z,T)=-k_{y}y\pm i\int \sqrt{m^{2}-\lambda ^{2}+\left( k_{y}+\omega
_{o}z\right) ^{2}}dz\pm i\frac{\lambda }{\omega }e^{\omega X}\pm \frac{%
\lambda }{\omega }e^{-\omega T}.  \label{k26}
\end{equation}
The wave function $u(X,y,z,T)=e^{iS}$ gives the quasi-classical asymptotes
of the solutions of the Klein-Gordon and Dirac equations. In the remote
past, as $T\longrightarrow -\infty ,$ $u_{-\infty }(X,y,z,T)$ takes the form

\begin{equation}
u_{-\infty }(X,y,z,T)=C(y,z)e^{-\frac{\lambda }{\omega }e^{\omega X}}c_{\mp
}e^{\mp i\frac{\lambda }{\omega }e^{-\omega T}}  \label{k27}
\end{equation}
where in (\ref{k27}) the upper and lower signs correspond respectively to
positive and negative frequency modes.

Analogously, we have that as $X\longrightarrow \infty $ and $%
T\longrightarrow \infty $ (\ref{k25}) reduces to

\begin{equation}
S(X,y,z,T)=-k_{y}y\pm i\int \sqrt{m^{2}-\lambda ^{2}+\left( k_{y}+\omega
_{o}z\right) ^{2}}dz\pm i\frac{\lambda }{\omega }e^{\omega X}\pm \sqrt{k}%
T\mp \frac{\lambda ^{2}e^{-2\omega T}}{2\omega \sqrt{k}}  \label{k28}
\end{equation}
and consequently,

\begin{equation}
u_{\infty }(X,y,z,T)=C^{\prime }(y,z)e^{-\frac{\lambda }{\omega }e^{\omega
X}}c_{\mp }^{\prime }e^{\mp i\sqrt{k}T}\exp (\mp \frac{i\lambda
^{2}e^{-2\omega T}}{2\omega \sqrt{k}})  \label{k29}
\end{equation}
where, in the present case, the upper sign corresponds to positive frequency
modes and the lower sign to negative modes. \ 

The results (\ref{k27}) and (\ref{k29}) give the quasi-classical asymptotic
behaviors of the relativistic wave equations in the accelerated coordinates (%
\ref{1}).

\section{Solution of the Klein Gordon equation}

In this section we proceed to solve the Klein-Gordon equation, coupled to a
constant magnetic field, in the accelerated coordinates with the line
element given by (\ref{2}).

The covariant generalization of the Klein Gordon equation is\cite
{Grib,Bagrov} 
\begin{equation}
g^{\alpha \beta }(\nabla _{\alpha }-ieA_{\alpha })(\nabla _{\beta
}-ieA_{\beta })\Phi -m^{2}\Phi =0  \label{KG}
\end{equation}
where $\nabla _{\alpha }=\partial _{\alpha }-\Gamma _{\alpha }$ is the
covariant derivative, and $A_{\alpha }$ is the vector potential given by (%
\ref{A}).

Substituting Eq. (\ref{2}) into (\ref{KG}) we readily obtain

\begin{equation}  \label{kgC}
\left[ \frac{1}{e^{-2\omega T}+e^{2\omega X}}\left( \partial
_{T}^{2}-\partial _{X}^{2}\right) -\left( \partial _{y}^{2}+\partial
_{z}^{2}\right) +i2eB_{x}z\partial _{y}+e^{2}B_{x}^{2}z^{2}-m^{2}\right]
\psi =0.  \label{k6}
\end{equation}
Since Eq. (\ref{k6}) commutes with the operator $-i\partial _{y}$ we can
look for a solution of the form $\psi =\phi (X,z,T)e^{ik_{y}y}$ which
reduces Eq. (\ref{k6}) to 
\begin{equation}
\left\{ \frac{1}{e^{-2\omega T}+e^{2\omega X}}\left( \partial
_{T}^{2}-\partial _{X}^{2}\right) +\left( eB_{x}z-k_{y}\right) ^{2}-\partial
_{z}^{2}-m^{2}\right\} \phi =0.  \label{k7}
\end{equation}
Eq. (\ref{k7}) can be separated in the form $\phi (X,z,T)=\eta (X,T)f(z)$.
The resulting equations are

\begin{equation}
\left[ \frac{1}{e^{-2\omega T}+e^{2\omega X}}\left( \partial
_{T}^{2}-\partial _{X}^{2}\right) +\lambda ^{2}\right] \eta =0  \label{k8}
\end{equation}
\begin{equation}
\frac{d^{2}f}{dz^{2}}=\left[ \left( \omega _{o}z-k_{y}\right) ^{2}+\left(
-m^{2}-\lambda ^{2}\right) \right] f(z)  \label{k9}
\end{equation}
where $\lambda $ is a separation constant. Variables $X$ and $T$ can be
separated in Eq. (\ref{k8}) after making the substitution: $\eta
(X,T)=f_{X}(X)f_{T}(T)$. The resulting equations for $X$ and $T$ are

\begin{equation}
\frac{d^{2}f_{X}}{dX^{2}}=-\left( -\lambda ^{2}e^{2\omega X}+\epsilon
^{2}\right) f_{X}(X)  \label{k11}
\end{equation}

\begin{equation}
\frac{d^{2}f_{T}}{dT^{2}}=-\left( +\lambda ^{2}e^{-2\omega T}+\epsilon
^{2}\right) f_{T}(T)  \label{k12}
\end{equation}
where $\epsilon $ is a constant of separation. Eq. (\ref{k11}) takes a more
familiar form in terms of Bessel functions\cite{abramowitz,Lebedev} after
introducing the variable $u=e^{\omega X}$

\begin{equation}
u^{2}\frac{d^{2}f_{X}}{du^{2}}+u\frac{df_{X}}{du}+\left( \frac{\epsilon ^{2}%
}{\omega ^{2}}-\frac{\lambda ^{2}}{\omega ^{2}}u^{2}\right) f_{u}=0
\label{k16}
\end{equation}
whose solution are the modified Bessel functions\cite{abramowitz} \ $I_{i\nu
}(z)$ and $K_{i\nu }(z)$%
\begin{equation}
f_{X}(X)=AI_{\pm i\nu }\left( \frac{\lambda }{\omega }e^{\omega X}\right)
+BK_{i\nu }\left( \frac{\lambda }{\omega }e^{\omega X}\right)
\end{equation}
where $A$ , $B$ are arbitrary constants and $\nu =\frac{\epsilon }{\omega }$%
. The solution of Eq. (\ref{k12}) can be obtained in the same manner.
Introducing the change of variables $v=e^{-\omega T}$ \ in (\ref{k12}) we
get the Bessel equation

\begin{equation}  \label{k19}
v^2\frac{d^2f_T}{dv^2}+v\frac{df_T}{dv}+\left( \frac k{\omega ^2}+ \frac{%
\lambda ^2}{\omega ^2}v^2\right) f_v=0
\end{equation}
whose solutions can be expressed in terms of the Hankel functions \cite
{abramowitz} $H_\nu ^{(1)}(z)$ and $H_\nu ^{(2)}(z)$

\begin{equation}
f_{T}(T)=A^{\prime }H_{i\nu }^{(1)}\left( \frac{\lambda }{\omega }e^{-\omega
T}\right) +B^{\prime }H_{i\nu }^{(2)}\left( \frac{\lambda }{\omega }%
e^{-\omega T}\right)  \label{k22}
\end{equation}
where $A^{\prime }$ and $B^{\prime }$ are arbitrary constants.

In order to solve Eq.(\ref{k9}), we introduce the new variable $x=\sqrt{%
\frac 2{\omega _o}}(\omega _oz+k_y)$. Eq. (\ref{k9}) takes the form

\begin{equation}
\frac{d^{2}f}{dx^{2}}-\left[ \frac{x^{2}}{4}+a\right] f(x)=0  \label{k14}
\end{equation}
where $a=-\frac{\mu ^{2}+\lambda ^{2}}{2\omega _{o}}$. Eq. (\ref{k14}) is
the Parabolic cylinder equation \cite{abramowitz}. Therefore the solution of
Eq. (\ref{k9} ) is 
\begin{equation}
f(x)=U\left( a,\sqrt{\frac{2}{\omega _{o}}}(\omega _{o}z+k_{y})\right)
\label{k15}
\end{equation}

Let us analyze the asymptotic behavior of the solutions of Eq. (\ref{kgC})
As $X\longrightarrow \infty $ and $T\longrightarrow -\infty $ we obtain that

\begin{equation}
\begin{array}{c}
\eta _{-\infty }(X,T)=K_{i\nu }\left( \frac{\lambda }{\omega }e^{\omega
X}\right) \left[ A_{-\infty }H_{i\nu }^{(1)}\left( \frac{\lambda }{\omega }%
e^{-\omega T}\right) +B_{-\infty }H_{i\nu }^{(2)}\left( \frac{\lambda }{%
\omega }e^{-\omega T}\right) \right] = \\ 
=e^{-\frac{\lambda }{\omega }e^{\omega X}}\left[ A_{-\infty }^{\prime }e^{i%
\frac{\lambda }{\omega }e^{-\omega T}}+B_{-\infty }^{\prime }e^{-i\frac{%
\lambda }{\omega }e^{-\omega T}}\right]
\end{array}
\label{k30}
\end{equation}
comparing (\ref{k30}) with (\ref{k27}) we identify the first right hand side
term as a positive frequency mode, and the second term as a negative
frequency. On the other hand when $X\longrightarrow \infty $ and $%
T\longrightarrow \infty $ we obtain

\begin{equation}
\begin{array}{c}
\eta _{\infty }(X,T)=K_{i\nu }\left( \frac{\lambda }{\omega }e^{\omega
X}\right) \left[ A_{\infty }J_{i\nu }\left( \frac{\lambda }{\omega }%
e^{-\omega T}\right) +B_{\infty }J_{-i\nu }\left( \frac{\lambda }{\omega }%
e^{-\omega T}\right) \right] = \\ 
=e^{-\frac{\lambda }{\omega }e^{\omega X}}\left[ A_{\infty }^{\prime }\left(
e^{-\omega T}\right) ^{i\nu }+B_{\infty }^{\prime }\left( e^{-\omega
T}\right) ^{-i\nu }\right] =e^{-\frac{\lambda }{\omega }e^{\omega X}}\left(
A_{\infty }^{\prime }e^{-i\sqrt{k}T}+B_{\infty }^{\prime }e^{i\sqrt{k}%
T}\right)
\end{array}
\label{k31}
\end{equation}

Also, comparing (\ref{k31}) with (\ref{k29}) we can identify the first and
second right hand side terms in (\ref{k31}) as positive and negative
frequency modes respectively.

Now, we are going to express an inertial positive frequency mode ($%
T\longrightarrow -\infty $) 
\begin{equation}
\eta _{inertial}(X,T)=C_{0}K_{i\nu }\left( \frac{\lambda }{\omega }e^{\omega
X}\right) H_{i\nu }^{(1)}\left( \frac{\lambda }{\omega }e^{-\omega T}\right)
\label{k32}
\end{equation}
in terms of the accelerated modes in the asymptotic future ($%
T\longrightarrow +\infty )$%
\begin{equation}
\eta _{acc}(X,T)=C_{1}K_{i\nu }\left( \frac{\lambda }{\omega }e^{\omega
X}\right) J_{i\nu }\left( \frac{\lambda }{\omega }e^{-\omega T}\right)
\end{equation}
where $C_{1}$ and $C_{0}$ are normalization constants according to the
standard inner product\cite{Birrel} $\ \left\langle \eta _{i},\eta
_{j}\right\rangle =-i\int (\eta _{i}\vec{\partial}_{s}\eta _{j}^{\ast }$ $%
-\eta _{j}\vec{\partial}_{s}\eta _{i}^{\ast })dS^{s}$ for the Klein Gordon
equation. The recurrence relation between $H_{i\nu }^{(1)}(z)$ and $J_{i\nu
}(z)$ permits one to express $\eta _{inertial}(X,T)$ in terms of $\eta
_{acc}(X,T)$ as follows

\begin{equation}
\eta _{inertial}(X,T)=\frac{C_{0}}{C_{1}}\left[ \frac{e^{\pi \nu }}{\sinh
\pi \nu }\eta _{acc}(X,T)-\frac{1}{\sinh \pi \nu }\eta _{acc}^{\ast }(X,T)%
\right] .  \label{relat}
\end{equation}
Since the inertial and accelerated modes are related via the Bogoliubov
coefficients\cite{Birrel,Grib} $\alpha $ and $\beta $ 
\begin{equation}
\eta _{i(inertial)}=\sum_{j}\alpha _{ij}\eta _{j(acc)}+\beta _{ij}\eta
_{j(acc)}^{\ast }
\end{equation}
we have that Eq. (\ref{relat}) gives immediately the values of $\alpha _{ij}$
and $\beta _{ij}$%
\begin{equation}
\alpha _{ij}=\frac{C_{0}}{C_{1}}\frac{e^{\pi \nu }}{\sinh \pi \nu }\delta
_{ij}=\alpha \delta _{ij},\quad \beta _{ij}=-\frac{C_{0}}{C_{1}}\frac{1}{%
\sinh \pi \nu }\delta _{ij}=\beta \delta _{ij}.  \label{coeff}
\end{equation}
Since $\left| \alpha \right| ^{2}-\left| \beta \right| ^{2}=1$ and 
\begin{equation}
\left| \frac{\beta }{\alpha }\right| =e^{-\pi \nu }
\end{equation}
the density of created particles\cite{Mishima} has the form

\begin{equation}
<0_{acc}|N|0_{acc}>=\left| \beta \right| ^{2}=\frac{1}{e^{2\pi \nu }-1}
\label{k32.4}
\end{equation}
which can be identified as a Planck distribution with a temperature

\begin{equation}
{\cal T}_{o}=\frac{\omega }{2\pi K_{B}}  \label{k35}
\end{equation}
The temperature measured by the accelerated observer can be obtained using
the relation \cite{Birrel,tolman}:

\begin{equation}
{\cal T}=\left( g_{00}\right) ^{-1/2}{\cal T}_{o}  \label{k36}
\end{equation}
In the asymptotic limit as $T\rightarrow +\infty $ we have $%
\lim_{T\rightarrow +\infty }\left( g_{00}\right) ^{-1/2}=e^{-\omega X}$ and
the temperature $T$ takes the value 
\begin{equation}
{\cal T}=\frac{\omega e^{-\omega X}}{2\pi K_{B}}=\frac{a_{+\infty }}{2\pi
K_{B}}
\end{equation}
then we obtain that the temperature is proportional to the asymptotic value
of the acceleration.

\section{Solution of the Dirac-Pauli equation}

In this section we solve the Dirac equation with anomalous magnetic moment
in the accelerated coordinates (\ref{1}) when a constant magnetic field is
present.

The covariant generalization of the Dirac-Pauli equation in curvilinear
coordinates is\cite{Bautista,Bagrov} 
\begin{equation}
\left\{ \gamma ^{\alpha }(\partial _{\alpha }-\Gamma _{\alpha })+\frac{\mu }{%
2}\gamma ^{\alpha }\gamma ^{\beta }F_{\alpha \beta }+m\right\} \Psi =0
\label{Dirac1}
\end{equation}
where $\gamma ^{\alpha }$ are the curvilinear Dirac matrices satisfying the
anticommutation relations $\left\{ \gamma ^{\alpha },\gamma ^{\beta
}\right\} _{+}=2g^{\alpha \beta }$, $\Gamma _{\alpha }$ are the spinor
connections and $F_{\alpha \beta }$ is the electromagnetic tensor. The
curvilinear $\gamma ^{\alpha }$ matrices are related to the constant
Minkowski $\tilde{\gamma}^{i}$ matrices with $\left\{ \tilde{\gamma}^{i},%
\tilde{\gamma}^{j}\right\} _{+}=2\eta ^{ij}$ via the tetrad $h_{\hspace{0.2cm%
}i}^{\alpha }$%
\begin{equation}
\gamma ^{\alpha }=h_{i}^{\alpha }\tilde{\gamma}^{i}
\end{equation}
In order to write the curvilinear Dirac matrices in Eq (\ref{Dirac1}) we
have to choose a tetrad $h_{\hspace{0.2cm}i}^{\alpha }.$ Here we are going
to work in the diagonal tetrad gauge, where $h_{\hspace{0.2cm}i}^{\alpha }$
takes the form

\begin{equation}
h_{\hspace{0.2cm}i}^{\alpha }=\left( 
\begin{array}{cccc}
\frac{1}{\sqrt{e^{-2\omega T}+e^{2\omega X}}} & 0 & 0 & 0 \\ 
0 & \frac{1}{\sqrt{e^{-2\omega T}+e^{2\omega X}}} & 0 & 0 \\ 
0 & 0 & 1 & 0 \\ 
0 & 0 & 0 & 1
\end{array}
\right) .  \label{diag}
\end{equation}
It is easy to verify that $g^{ik}=\eta ^{jl}h_{\hspace{0.2cm}j}^{i}h_{%
\hspace{0.2cm}l}^{k}$ . In this tetrad gauge, the curvilinear Dirac matrices
can be expressed in terms of the constant $\tilde{\gamma}^{i}$ as follows

\begin{equation}
\begin{array}{c}
\gamma ^{0}=\frac{\overline{\tilde{\gamma}}^{0}}{\sqrt{e^{-2\omega
T}+e^{2\omega X}}},\quad \gamma ^{1}=\frac{\overline{\tilde{\gamma}}^{1}}{%
\sqrt{e^{-2\omega T}+e^{2\omega X}}},\quad \gamma ^{2}=\overline{\tilde{%
\gamma}}^{2},\quad \gamma ^{3}=\overline{\tilde{\gamma}}^{3}.
\end{array}
\label{rel7}
\end{equation}
In the Diagonal tetrad gauge (\ref{diag}) the spinor connections, defined by
the relation\cite{Brill}: 
\begin{equation}
\Gamma _{\mu }=\frac{1}{4}g_{\lambda \alpha }\left( \frac{\partial h_{\nu
}^{.i}}{\partial x^{\mu }}h_{.i}^{\alpha }-\Gamma _{\nu \mu }^{\alpha
}\right) s^{\lambda \nu }  \label{conexion}
\end{equation}
with $s^{\lambda \nu }=\frac{1}{2}\left( \gamma ^{\lambda }\gamma ^{\nu
}-\gamma ^{\nu }\gamma ^{\lambda }\right) $, take the form 
\begin{equation}
\begin{array}{c}
\Gamma _{0}=-\frac{1}{2}\omega e^{2\omega X}\left( e^{-2\omega T}+e^{2\omega
X}\right) ^{-1}\overline{\gamma }^{1}\overline{\gamma }^{0},\quad \Gamma
_{1}=\frac{1}{2}\omega e^{-2\omega T}\left( e^{-2\omega T}+e^{2\omega
X}\right) ^{-1}\overline{\gamma }^{1}\overline{\gamma }^{0}, \\ 
\Gamma _{2}=0,\quad \Gamma _{3}=0.
\end{array}
\label{gammas}
\end{equation}
Substituting (\ref{gammas}) , (\ref{rel7}) and (\ref{A}) into (\ref{Dirac1})
we obtain

\begin{equation}
\begin{array}{c}
\frac{\overline{\gamma }^{1}}{\sqrt{e^{2\omega X}+e^{-2\omega T}}}\left[ 
\frac{\partial \Psi }{\partial X}+\frac{\omega e^{2\omega X}}{2\left(
e^{2\omega X}+e^{-2\omega T}\right) }\Psi \right] +\overline{\gamma }%
^{2}\left( \frac{\partial \Psi }{\partial y}-i\frac{eB_{x}}{\sqrt{2}}z\Psi
\right) +\overline{\gamma }^{3}\frac{\partial \Psi }{\partial z}+ \\ 
+\frac{\overline{\gamma }^{4}}{\sqrt{e^{2\omega X}+e^{-2\omega T}}}\left[ 
\frac{\partial \Psi }{\partial T}-\frac{\omega e^{-2\omega T}}{2\left(
e^{2\omega X}+e^{-2\omega T}\right) }\Psi \right] +m\Psi +i\mu B_{x}%
\overline{\gamma }^{2}\overline{\gamma }^{3}\Psi =0.
\end{array}
\end{equation}
Introducing the spinor $\Phi ,$

\begin{equation}
\Psi =\frac{\Phi }{\left( e^{2\omega X}+e^{-2\omega T}\right) ^{1/4}}
\label{c2}
\end{equation}
we eliminate the contribution terms due to the spinor connections. The Dirac
equation takes the form

\begin{equation}
\frac{\overline{\gamma }^{1}\frac{\partial \Phi }{\partial X}+\overline{%
\gamma }^{4}\frac{\partial \Phi }{\partial T}}{\sqrt{e^{2\omega
X}+e^{-2\omega T}}}+\overline{\gamma }^{2}\left( \frac{\partial \Phi }{%
\partial y}-ieB_{x}z\Phi \right) +\overline{\gamma }^{3}\frac{\partial \Phi 
}{\partial z}+m\Phi +i\mu _{o}\overline{\gamma }^{2}\overline{\gamma }%
^{3}\Phi =0  \label{Dirac2}
\end{equation}
where $\mu _{o}=\mu B_{x}$ . In order to solve Eq. (\ref{Dirac2}) we proceed
to separate variables using the algebraic method of separation developed by
Shishkin and Villalba\cite{Villalba,Shishkin,Shishkin2,Shishkin3} The idea
behind the method is to reduce Eq. (\ref{Dirac2}) to a sum of two commuting
first order differential operators 
\begin{equation}
\left( \hat{K}_{1}(T,X)+\hat{K}_{2}(y,z)\right) \psi =0,\quad \left[ \hat{K}%
_{1}(T,X),\hat{K}_{2}(y,z)\right] _{-}=0  \label{K}
\end{equation}
where in the present case, $\psi =\overline{\gamma }^{3}\overline{\gamma }%
^{2}\Phi $ and 
\begin{equation}
\hat{K}_{1}=\frac{\overline{\gamma }^{1}\overline{\gamma }^{2}\overline{%
\gamma }^{3}\partial _{X}+\overline{\gamma }^{4}\overline{\gamma }^{2}%
\overline{\gamma }^{3}\partial _{T}}{\sqrt{e^{2\omega X}+e^{-2\omega T}}}%
,\quad \hat{K}_{2}=\overline{\gamma }^{3}\left( \partial _{y}-i\frac{eB_{x}}{%
\sqrt{2}}z\right) -\overline{\gamma }^{2}\partial _{z}+\overline{\gamma }^{2}%
\overline{\gamma }^{3}m-i\mu _{o}.  \label{K2}
\end{equation}
Since $\hat{K}_{1}$ and $\hat{K}_{2}$ satisfy (\ref{K}) they satisfy the
eigenvalue equations 
\begin{equation}
K_{1}\psi =-i\lambda \psi ,\quad K_{2}\psi =i\lambda \psi .  \label{c7}
\end{equation}
Now, we proceed to solve equation $K_{1}\psi =-i\lambda \psi :$

\begin{equation}
\left( \overline{\gamma }^{1}\overline{\gamma }^{2}\overline{\gamma }%
^{3}\partial _{X}+\overline{\gamma }^{4}\overline{\gamma }^{2}\overline{%
\gamma }^{3}\partial _{T}\right) \psi =-i\lambda \sqrt{e^{2\omega
X}+e^{-2\omega T}}\psi  \label{c8}
\end{equation}
applying the transformation $\psi =S\phi $ defined by :

\begin{equation}
S=e^{\Xi (X,T)}e^{\left( i\overline{\gamma }^{1}\overline{\gamma }^{4}\Theta
(X,T)\right) }  \label{c9}
\end{equation}
we reduce Eq. (\ref{c8}) to the form

\begin{equation}
\begin{array}{c}
\left( \overline{\gamma }^{1}\overline{\gamma }^{2}\overline{\gamma }%
^{3}\partial _{X}+\overline{\gamma }^{4}\overline{\gamma }^{2}\overline{%
\gamma }^{3}\partial _{T}\right) \phi +\overline{\gamma }^{1}\overline{%
\gamma }^{2}\overline{\gamma }^{3}\left[ \partial _{X}\Xi (X,T)+i\partial
_{T}\Theta (X,T)\right] \phi + \\ 
+\overline{\gamma }^{4}\overline{\gamma }^{2}\overline{\gamma }^{3}\left[
\partial _{T}\Xi (X,T)+i\partial _{X}\Theta (X,T)\right] \phi =-i\lambda 
\sqrt{e^{2\omega X}+e^{-2\omega T}}e^{2i\overline{\gamma }^{1}\overline{%
\gamma }^{4}\Theta (X,T)}\phi
\end{array}
\label{c10}
\end{equation}
In order to separate variables in (\ref{c10}) we demand that the terms
inside the brackets vanish, i.e. 
\begin{equation}
\partial _{X}\Xi (X,T)=-i\partial _{T}\Theta (X,T),\quad \partial _{T}\Xi
(X,T)=-i\partial _{X}\Theta (X,T)  \label{c11}
\end{equation}
the solution of Eq. (\ref{c11}) is 
\begin{equation}
\Xi (X,T)=\frac{i}{2}\arctan \left( e^{\omega (X+T)}\right) ,\quad \Theta
(X,T)=\frac{1}{2}\arctan \left( e^{-\omega (X+T)}\right)  \label{c12}
\end{equation}
which determines the spinor transformation S Eq. (\ref{c9}). Eq. (\ref{c10})
reduces to

\begin{equation}
\left( \overline{\gamma }^{1}\overline{\gamma }^{2}\overline{\gamma }%
^{3}\partial _{X}+\overline{\gamma }^{4}\overline{\gamma }^{2}\overline{%
\gamma }^{3}\partial _{T}\right) \phi =i\lambda \left( e^{\omega X}+i%
\overline{\gamma }^{1}\overline{\gamma }^{4}e^{-\omega T}\right) \phi .
\label{c13}
\end{equation}
Let $L_{1}$ and $L_{2}$ be the commuting operators 
\begin{equation}
L_{1}=\overline{\gamma }^{1}\overline{\gamma }^{2}\overline{\gamma }%
^{3}\partial _{X}-i\lambda e^{\omega X},\quad L_{2}=\overline{\gamma }^{1}%
\overline{\gamma }^{2}\overline{\gamma }^{3}\partial _{T}+\lambda e^{-\omega
T},\quad \left[ L_{1},L_{2}\right] =0.
\end{equation}
Eq. (\ref{c13}) can be expressed in terms of \ $L_{1}$ and $L_{2}$ as
follows:

\begin{equation}
\left( L_{1}+L_{2}\overline{\gamma }^{1}\overline{\gamma }^{4}\right) \phi
=0.  \label{c16}
\end{equation}
In order to solve Eq. (\ref{c16}) we introduce the auxiliary spinor

\begin{equation}
\phi =\left( \overline{\gamma }^{1}\overline{\gamma }^{4}L_{1}-L_{2}\right)
W,  \label{c17}
\end{equation}
substituting (\ref{c17}) into (\ref{c16}) we have that $\left( L_{1}%
\overline{\gamma }^{1}\overline{\gamma }^{4}L_{1}-L_{2}\overline{\gamma }^{1}%
\overline{\gamma }^{4}L_{2}\right) W=0.$ This allows separation of variables
as follows:

\begin{equation}
\left( \partial _{X}^{2}-\lambda ^{2}e^{2\omega X}+i\lambda \omega e^{\omega
X}\overline{\gamma }^{1}\overline{\gamma }^{2}\overline{\gamma }^{3}\right)
W=-\epsilon ^{2}W  \label{c20}
\end{equation}

\begin{equation}
\left( \partial _{T}^{2}+\lambda ^{2}e^{-2\omega T}+\lambda \omega
e^{-\omega T}\overline{\gamma }^{1}\overline{\gamma }^{2}\overline{\gamma }%
^{3}\right) W=-\epsilon ^{2}W  \label{c21}
\end{equation}
where $\epsilon $ is a constant of separation.

When we choose the following representation for the Dirac matrices\cite
{Jauch} $\widetilde{\gamma }^{i}$:

\begin{equation}
\begin{array}{cc}
\tilde{\gamma}^{1}=\left( 
\begin{array}{cc}
0 & \sigma _{1} \\ 
\sigma _{1} & 0
\end{array}
\right) & \tilde{\gamma}^{2}=\left( 
\begin{array}{cc}
0 & \sigma _{2} \\ 
\sigma _{2} & 0
\end{array}
\right) ,\quad 
\begin{array}{cc}
\tilde{\gamma}^{3}=\left( 
\begin{array}{cc}
1 & 0 \\ 
0 & -1
\end{array}
\right) & \tilde{\gamma}^{4}=\left( 
\begin{array}{cc}
0 & i\sigma _{3} \\ 
i\sigma _{3} & 0
\end{array}
\right)
\end{array}
\end{array}
\label{c22}
\end{equation}
then the spinor $W$ has the structure:

\begin{equation}
W=\left( 
\begin{array}{c}
\alpha (X)A(T) \\ 
\beta (X)B(T) \\ 
\gamma (X)C(T) \\ 
\delta (X)D(T)
\end{array}
\right) .  \label{c23}
\end{equation}
Substituting \ (\ref{c23}) into Eq. (\ref{c20}) we obtain 
\begin{equation}
\begin{array}{c}
\left( d_{X}^{2}-\lambda ^{2}e^{2\omega X}-\lambda \omega e^{\omega
X}+\epsilon ^{2}\right) \left( 
\begin{array}{c}
\alpha (X) \\ 
\delta (X)
\end{array}
\right) =0, \\ 
\left( d_{X}^{2}-\lambda ^{2}e^{2\omega X}+\lambda \omega e^{\omega
X}+\epsilon ^{2}\right) \left( 
\begin{array}{c}
\beta (X) \\ 
\gamma (X)
\end{array}
\right) =0;
\end{array}
\label{c24}
\end{equation}
analogously, Eq. (\ref{c21}) reduces to 
\begin{equation}
\begin{array}{c}
\left( \partial _{T}^{2}+\lambda ^{2}e^{-2\omega T}+i\lambda \omega
e^{-\omega T}+\epsilon ^{2}\right) \left( 
\begin{array}{c}
A(T) \\ 
D(t)
\end{array}
\right) =0, \\ 
\left( \partial _{T}^{2}+\lambda ^{2}e^{-2\omega T}-i\lambda \omega
e^{-\omega T}+\epsilon ^{2}\right) \left( 
\begin{array}{c}
B(T) \\ 
C(t)
\end{array}
\right) =0
\end{array}
\label{c25}
\end{equation}
In order to solve the system of equations (\ref{c24}), it suffices to solve
the second order equation

\begin{equation}
u^{2}\frac{d^{2}f}{du^{2}}+u\frac{df}{du}+\left( -\frac{\lambda ^{2}}{\omega
^{2}}u^{2}\pm \frac{\lambda }{\omega }u+\frac{\epsilon ^{2}}{{\omega ^{2}}}%
\right) f=0  \label{c27}
\end{equation}
where $u=e^{\omega X}.$

Analogously, we have that solving the system of equations (\ref{c25}) is
equivalent to solve the Whittaker\cite{abramowitz} differential equation

\begin{equation}
v^{2}\frac{d^{2}g}{dv^{2}}+v\frac{dg}{dv}+\left( \frac{\lambda ^{2}}{\omega
^{2}}v^{2}\pm \frac{i\lambda }{\omega }v+\frac{\epsilon ^{2}}{{\omega ^{2}}}%
\right) g=0  \label{c29}
\end{equation}
where we have introduced the change of variable $v=e^{-\omega T}.$ The
solution of Eqs. (\ref{c27}) and (\ref{c29}) can be expressed in terms of a
combination of Whittaker functions

\begin{equation}
\begin{array}{c}
\digamma _{\kappa ,\mu }(z)=C_{1}M_{\kappa ,\mu }(z)+C_{2}W_{\kappa ,\mu
}(z)= \\ 
C_{1}e^{\frac{-z}{2}}z^{\frac{1}{2}+\mu }M\left( \frac{1}{2}+\mu -\kappa
,1+2\mu ;z\right) +C_{2}e^{\frac{-z}{2}}z^{\frac{1}{2}+\mu }U\left( \frac{1}{%
2}+\mu -\kappa ,1+2\mu ;z\right) .
\end{array}
\label{c31}
\end{equation}
The solutions of \ Eqns. (\ref{c27}) \ and (\ref{c29}) are 
\begin{equation}
f(u)=\frac{\digamma _{\pm 1/2,\mu }\left( \frac{2\lambda }{\omega }u\right) 
}{\sqrt{u}},\quad g(v)=\frac{\digamma _{\pm 1/2,\mu }\left( \frac{2i\lambda 
}{\omega }v\right) }{\sqrt{v}}
\end{equation}
where $\mu =\frac{i\epsilon }{\omega }.$

The spinor $\phi $ can be computed with the help of the inverse
transformation (\ref{c17}):

\begin{equation}
\phi =\left( \overline{\gamma }^{1}\overline{\gamma }^{4}L_{1}-L_{2}\right)
W=\left( 
\begin{array}{c}
-i\alpha (X)\left( \partial _{T}-i\lambda e^{-\omega T}\right)
A(T)-B(T)\left( \partial _{X}+\lambda e^{\omega X}\right) \beta (X) \\ 
i\beta (X)\left( \partial _{T}+i\lambda e^{-\omega T}\right) B(T)-A(T)\left(
\partial _{X}-\lambda e^{\omega X}\right) \alpha (X) \\ 
i\gamma (X)\left( \partial _{T}+i\lambda e^{-\omega T}\right)
C(T)+D(T)\left( \partial _{X}-\lambda e^{\omega X}\right) \delta (X) \\ 
-i\delta (X)\left( \partial _{T}-i\lambda e^{-\omega T}\right)
D(T)+C(T)\left( \partial _{X}+\lambda e^{\omega X}\right) \gamma (X)
\end{array}
\right) .  \label{c34}
\end{equation}
The system of equations (\ref{c24}) can be written as a coupled system of
ordinary differential equations

\begin{equation}
\begin{array}{c}
\left( \partial _{X}-\lambda e^{\omega X}\right) \alpha (X)=i\epsilon \beta
(X),\quad \left( \partial _{X}+\lambda e^{\omega X}\right) \beta
(X)=i\epsilon \alpha (X), \\ 
\left( \partial _{X}+\lambda e^{\omega X}\right) \gamma (X)=\epsilon \delta
(X),\quad \left( \partial _{X}-\lambda e^{\omega X}\right) \delta
(X)=-\epsilon \gamma (X).
\end{array}
\label{c35}
\end{equation}
analogously, we have that the Eq. (\ref{c25}) is equivalent to

\begin{equation}
\begin{array}{c}
\left( \partial _{T}-i\lambda e^{-\omega T}\right) A(T)=i\epsilon B(T),\quad
\left( \partial _{T}+i\lambda e^{-\omega T}\right) B(T)=i\epsilon A(T) \\ 
\left( \partial _{T}+i\lambda e^{-\omega T}\right) C(T)=\epsilon D(T),\quad
\left( \partial _{T}-i\lambda e^{-\omega T}\right) D(T)=-\epsilon C(T).
\end{array}
\quad  \label{c36}
\end{equation}
Substituting (\ref{c35}) and (\ref{c36}) into (\ref{c34}) we arrive at

\begin{equation}
\phi =\epsilon \left( 
\begin{array}{c}
(-i+1)\alpha (x)B(T) \\ 
(-i-1)\beta (x)A(T) \\ 
(i-1)\gamma (x)D(T) \\ 
(i+1)\delta (x)C(T)
\end{array}
\right)  \label{c37}
\end{equation}
where 
\begin{equation}
\begin{array}{c}
\alpha (u)=\frac{c_{1}}{\sqrt{u}}M_{-1/2,\mu }\left( \frac{2\lambda }{\omega 
}u\right) ,\ \beta (u)=\frac{c_{2}}{\sqrt{u}}M_{1/2,\mu }\left( \frac{%
2\lambda }{\omega }u\right) , \\ 
\gamma (u)=\frac{c_{3}}{\sqrt{u}}M_{1/2,\mu }\left( \frac{2\lambda }{\omega }%
u\right) ,\ \delta (u)=\frac{c_{4}}{\sqrt{u}}M_{-1/2,\mu }\left( \frac{%
2\lambda }{\omega }u\right)
\end{array}
\end{equation}
and 
\begin{equation}
\begin{array}{c}
A(T)=\frac{d_{1}}{\sqrt{v}}\digamma _{1/2,\mu }\left( \frac{2i\lambda }{%
\omega }v\right) ,\ B(T)=\frac{d_{2}}{\sqrt{v}}\digamma _{-1/2,\mu }\left( 
\frac{2i\lambda }{\omega }v\right) , \\ 
C(T)=\frac{d_{3}}{\sqrt{v}}\digamma _{-1/2,\mu }\left( \frac{2i\lambda }{%
\omega }v\right) ,\ \ D(T)=\frac{d_{4}}{\sqrt{v}}\digamma _{1/2,\mu }\left( 
\frac{2i\lambda }{\omega }v\right) .
\end{array}
\end{equation}

Using the recurrence relations for the Whittaker functions we find that $%
\phi $ has the form 
\begin{equation}
\phi =\left( 
\begin{array}{c}
a_{1}(i-1)f_{1} \\ 
-a_{1}(i+1)f_{2} \\ 
a_{2}(i+1)f_{2} \\ 
a_{2}(i-1)f_{1}
\end{array}
\right)  \label{fi}
\end{equation}
where 
\begin{equation}
f_{1}=\frac{\epsilon }{\sqrt{uv}}M_{-1/2,\mu }\left( \frac{2\lambda }{\omega 
}u\right) \digamma _{-1/2,\mu }\left( \frac{2i\lambda }{\omega }v\right) ,\
f_{2}=\frac{\epsilon }{\sqrt{u\nu }}M_{1/2,\mu }\left( \frac{2\lambda }{%
\omega }u\right) \digamma _{1/2,\mu }\left( \frac{2i\lambda }{\omega }%
v\right) .
\end{equation}
Recalling that $\psi =S\phi $ where $S$ is given by (\ref{c9}) we have that $%
\psi $ takes the form

\begin{equation}
\psi =\left( 
\begin{array}{c}
a_{1}\cos \Theta (i-1)f_{1}-a_{1}\sin \Theta (i+1)f_{2} \\ 
-a_{1}\sin \Theta (i-1)f_{1}-a_{1}\cos \Theta (i+1)f_{2} \\ 
a_{2}\cos \Theta (i+1)f_{2}+a_{2}\sin \Theta (i-1)f_{1} \\ 
-a_{2}\sin \Theta (i+1)f_{2}+a_{2}\cos \Theta (i-1)f_{1}
\end{array}
\right) e^{\Xi }.  \label{c52}
\end{equation}
Now we proceed to solve the equations (\ref{c7}) governing the dependence of
the spinor solution of the Dirac equation on the coordinates $y$ and $z$

\begin{equation}
\left( \overline{\gamma }^{3}\left( \partial _{y}-ieB_{x}z\right) -\overline{%
\gamma }^{2}\partial _{z}+\overline{\gamma }^{2}\overline{\gamma }^{3}m-i\mu
_{o}\right) \psi =i\lambda \psi .  \label{c53}
\end{equation}
Since $\left[ \hat{K}_{2},-i\partial _{y}\right] _{-}=0$, we have that $\psi 
$ can be written as

\begin{equation}
\psi =e^{ik_{y}y}\varphi (z),  \label{c54}
\end{equation}
substituting (\ref{c54}) into (\ref{c53}) we obtain

\begin{equation}
\left( -i\overline{\gamma }^{3}\left( -k_{y}+\omega _{o}z\right) -\overline{%
\gamma }^{2}\partial _{z}+\overline{\gamma }^{2}\overline{\gamma }^{3}\mu
\right) \varphi =i\left( \lambda +\mu _{o}\right) \varphi   \label{c55}
\end{equation}
where $\omega _{o}=eB_{x}$. Introducing the spinor $\varphi =\Sigma \theta $
with 
\begin{equation}
\Sigma =\frac{1-\overline{\gamma }^{1}\overline{\gamma }^{3}}{\sqrt{2}}
\label{c58}
\end{equation}
we find that Eq (\ref{c55}) reduces to

\begin{equation}
\left( -i\overline{\gamma }^{1}\left( -k_{y}+\omega _{o}z\right) -\overline{%
\gamma }^{2}\partial _{z}+\overline{\gamma }^{2}\overline{\gamma }^{1}\mu
\right) \theta =i\left( \lambda +\mu _{o}\right) \theta ;  \label{c59}
\end{equation}
substituting into (\ref{c59}) the Dirac matrices in the representation (\ref
{c22}) and considering a spinor with the structure 
\begin{equation}
\theta =\left( 
\begin{array}{c}
\theta _{1} \\ 
\theta _{2} \\ 
\theta _{3} \\ 
\theta _{4}
\end{array}
\right)  \label{c60}
\end{equation}
we reduce our problem to that of solving the following system of partial
differential equations: 
\begin{equation}
\left[ \partial _{z}-\left( -k_{y}+\omega _{o}z\right) \right] \theta
_{4}=\left( \mu +\lambda +\mu _{o}\right) \theta _{1},\ \left[ \partial
_{z}+\left( -k_{y}+\omega _{o}z\right) \right] \theta _{1}=\left( \mu
-\lambda -\mu _{o}\right) \theta _{4},  \label{uno}
\end{equation}
\begin{equation}
\left[ \partial _{z}-\left( -k_{y}+\omega _{o}z\right) \right] \theta
_{2}=\left( \mu +\lambda +\mu _{o}\right) \theta _{3},\ \left[ \partial
_{z}+\left( -k_{y}+\omega _{o}z\right) \right] \theta _{3}=\left( \mu
-\lambda -\mu _{o}\right) \theta _{2}.  \label{dos}
\end{equation}
Looking at (\ref{uno}) and (\ref{dos}) we see that $\theta _{1}\sim \theta
_{3}$ and $\theta _{2}\sim \theta _{4}$ and therefore it is only necessary
solve one of the systems of coupled equations. Making the change of variable 
$x=\sqrt{\frac{2}{\omega _{o}}}(\omega _{o}z-k_{y})$ we have

\begin{equation}
\frac{d^{2}}{dx^{2}}\theta _{1}-\left[ \frac{x^{2}}{4}+a\right] \theta
_{1}=0,  \label{c65}
\end{equation}

\begin{equation}
\frac{d^{2}}{dx^{2}}\theta _{4}-\left[ \frac{x^{2}}{4}+(a+1)\right] \theta
_{4}=0,  \label{c66}
\end{equation}
where $a=\frac{\mu ^{2}-\lambda ^{2}}{2\omega _{o}}-\frac{1}{2}$ . Equations
(\ref{c65}) and (\ref{c66}) are Parabolic cylinder equations \cite
{abramowitz} and their solutions are

\begin{equation}
\begin{array}{c}
\theta _{1}=d_{1}U\left( a,\sqrt{\frac{2}{\omega _{o}}(\omega _{o}z-k_{y})}%
\right) ,\ \theta _{3}=d_{3}U\left( a,\sqrt{\frac{2}{\omega _{o}}(\omega
_{o}z-k_{y})}\right) \\ 
\theta _{4}=d_{4}U\left( a+1,\sqrt{\frac{2}{\omega _{o}}}(\omega
_{o}z-k_{y})\right) ,\ \theta _{2}=d_{2}U\left( a+1,\sqrt{\frac{2}{\omega
_{o}}}(\omega _{o}z-k_{y})\right)
\end{array}
\label{c67}
\end{equation}
with the help of the recurrence relations for the parabolic cylinder
equation and the equations (\ref{uno}),(\ref{dos}) , we can find the
relation between the coefficients $d_{i}$:

\begin{equation}
\frac{d_{1}}{d_{4}}=-\frac{\sqrt{2\omega _{o}}}{\mu +\lambda +\mu _{o}}
\label{c71}
\end{equation}
and

\begin{equation}
\frac{d_{3}}{d_{2}}=-\frac{\sqrt{2\omega _{o}}}{\mu +\lambda +\mu _{o}},
\label{c72}
\end{equation}
consequently

\begin{equation}
\theta =\left( 
\begin{array}{c}
-\frac{\sqrt{2\omega _{o}}}{\mu +\lambda +\mu _{o}}d_{4}U(a,x) \\ 
d_{2}U(a+1,x) \\ 
-\frac{\sqrt{2\omega _{o}}}{\mu +\lambda +\mu _{o}}d_{2}U(a,x) \\ 
d_{4}U(a+1,x)
\end{array}
\right) .  \label{c73}
\end{equation}
The spinor $\varphi $ is obtained using the matrix transformation $\Sigma $ (%
\ref{c58}):

\begin{equation}
\varphi =\Sigma \theta =\frac{1}{\sqrt{2}}\left( 
\begin{array}{c}
\theta _{1}+\theta _{4} \\ 
\theta _{2}+\theta _{3} \\ 
-\theta _{2}+\theta _{3} \\ 
-\theta _{1}+\theta _{4}
\end{array}
\right) .  \label{c74}
\end{equation}
From (\ref{c58}) and (\ref{c54}) we obtain

\begin{equation}
\psi =\frac{1}{\sqrt{2}}\left( 
\begin{array}{c}
d_{4}\left( U(a+1,x)-\frac{\sqrt{2\omega _{o}}}{\mu +\lambda +\mu _{o}}%
U(a,x)\right) \\ 
d_{2}\left( U(a+1,x)-\frac{\sqrt{2\omega _{o}}}{\mu +\lambda +\mu _{o}}%
U(a,x)\right) \\ 
-d_{2}\left( U(a+1,x)+\frac{\sqrt{2\omega _{o}}}{\mu +\lambda +\mu _{o}}%
U(a,x)\right) \\ 
d_{4}\left( U(a+1,x)+\frac{\sqrt{2\omega _{o}}}{\mu +\lambda +\mu _{o}}%
U(a,x)\right)
\end{array}
\right) e^{ik_{y}y}.  \label{c75}
\end{equation}

Combining (\ref{c52}) with (\ref{c75}) we find that the spinor $\psi $ is

\begin{equation}
\psi =\frac{1}{\sqrt{2}}\left( 
\begin{array}{c}
\left( \cos \Theta (i-1)f_{1}-\sin \Theta (i+1)f_{2}\right) \left( U(a+1,x)-%
\frac{\sqrt{2\omega _{o}}}{\mu +\lambda +\mu _{o}}U(a,x)\right) \\ 
\left( -\sin \Theta (i-1)f_{1}-\cos \Theta (i+1)f_{2}\right) \left( U(a+1,x)-%
\frac{\sqrt{2\omega _{o}}}{\mu +\lambda +\mu _{o}}U(a,x)\right) \\ 
\left( \cos \Theta (i+1)f_{2}+\sin \Theta (i-1)f_{1}\right) \left( U(a+1,x)+%
\frac{\sqrt{2\omega _{o}}}{\mu +\lambda +\mu _{o}}U(a,x)\right) \\ 
\left( -\sin \Theta (i+1)f_{2}+\cos \Theta (i-1)f_{1}\right) \left( U(a+1,x)+%
\frac{\sqrt{2\omega _{o}}}{\mu +\lambda +\mu _{o}}U(a,x)\right)
\end{array}
\right) e^{ik_{y}y}e^{\Xi }  \label{c78}
\end{equation}

Now, we proceed to analyze the asymptotic limits as $T\rightarrow -\infty $
and $T\rightarrow +\infty $. We will confine our attention to the solutions
of the spinor (\ref{c78}). In the asymptotes we obtain a time dependent term
multiplied by a factor depending on space variables. Here, we proceed like
we did in Sec. III for the scalar case.

The relation between \ $W_{\lambda ,\mu }(z)$ and $M_{\lambda ,\mu }(z)$\cite
{Gradshtein} will be helpful

\begin{equation}
W_{\lambda ,\mu }(z)=\frac{\Gamma (-2\mu )}{\Gamma (\frac{1}{2}-\mu -\lambda
)}M_{\lambda ,\mu }(z)+\frac{\Gamma (2\mu )}{\Gamma (\frac{1}{2}+\mu
-\lambda )}M_{\lambda ,-\mu }(z)  \label{c79}
\end{equation}
Taking into account expression (\ref{c79}) we find that the solutions of Eq.
(\ref{c36}) are related as follows 
\begin{equation}
W_{\lambda ,\mu }(-\frac{2i\lambda }{\omega }v)=\frac{\Gamma (-2\mu )}{%
\Gamma (\frac{1}{2}-\mu -\lambda )}M_{\lambda ,\mu }(-\frac{2i\lambda }{%
\omega }v)+\frac{\Gamma (2\mu )}{\Gamma (\frac{1}{2}+\mu -\lambda )}e^{i\pi
(\mu -\frac{1}{2})}M_{-\lambda ,-\mu }(\frac{2i\lambda }{\omega }v)
\label{rec}
\end{equation}
Looking at the quasi-classical behavior given by Eq. (\ref{k29}), we
identify the positive and negative accelerated modes as

\begin{equation}
\psi _{+}^{acc}=N_{1}M_{\lambda ,\mu }(-\frac{2i\lambda }{\omega }v),\quad
\psi _{-}^{acc}=N_{1}M_{-\lambda ,-\mu }(\frac{2i\lambda }{\omega }v)
\end{equation}
where $N_{1}$ is a normalization constant. Also an inertial $%
(T\longrightarrow -\infty )$ positive frequency mode $\ \psi _{+}^{ine}$ is
given by 
\begin{equation}
\psi _{+}^{ine}=N_{3}W_{\lambda ,\mu }(-\frac{2i\lambda }{\omega }v)
\end{equation}
where $N_{3}$ is a normalization constant.

Looking at \ Eq. (\ref{rec}) and recalling that the inertial modes can be
expressed in terms of the accelerated positive and negative modes via the
Bogoliubov coefficients

\begin{equation}
\psi _{+}^{ine}=\alpha \psi _{+}^{acc}+\beta \psi _{-}^{acc},  \label{c87}
\end{equation}
we get that 
\begin{equation}
\left| \frac{\beta }{\alpha }\right| =e^{-\frac{\epsilon \pi }{\omega }}
\end{equation}
and taking into account that $\left| \alpha \right| ^{2}+\left| \beta
\right| ^{2}=1$, we find that the density of particles created is\cite
{Mishima}

\begin{equation}
<0_{acc}|N|0_{acc}>=\left| \beta \right| ^{2}=\frac{1}{1+e^{2\pi \frac{%
\epsilon }{\omega }}},  \label{c87.2}
\end{equation}
a result that can be identified as a Fermi-Dirac distribution of particles
associated with a temperature

\begin{equation}
{\cal T}_{o}=\frac{\omega }{2\pi K_{B}}  \label{c90}
\end{equation}
and, consequently, the temperature detected in the accelerated frame is\cite
{tolman}

\begin{equation}
{\cal T}=\left( g_{00}\right) ^{-1/2}{\cal T}_{o}  \label{c91}
\end{equation}
that in the asymptotic limit as $T\rightarrow +\infty $ takes the form 
\[
{\cal T}=\frac{\omega e^{-\omega X}}{4\pi K_{B}}, 
\]
showing that the temperature is proportional to the asymptotic value of the
acceleration.

\section{Discussion of the results}

In the present paper we have separated variables and solved the Klein-Gordon
and Dirac equations in the curvilinear coordinates (\ref{1}) when a constant
magnetic field (\ref{A}) is present. The algebraic method of separation\ 
\cite{Villalba,Shishkin,Shishkin2,Shishkin3} has been applied to reduce
Dirac-Pauli equation to a system of coupled ordinary differential equations.
Using the obtained exact solutions we calculated the density of scalar and
spin $1/2$ particles detected by an accelerated observer associated with
system of coordinates (\ref{1}) when a constant magnetic field in the
direction of acceleration is present. The identification of positive and
negative frequency modes was carried out comparing the relativistic
solutions with the quasi-classical \ Hamilton-Jacobi solutions (\ref{k27})
and (\ref{k29}). The results obtained in Sec III and IV indicate that the
magnetic field does not modify the thermal character of spectrum. The
temperature associated with the thermal bath is not modified by $B$. This
result has a classical counterpart: Since $B$ is colinear to the motion, it
does not accelerate the particle and no radiation is caused by it. The
presence of a non minimal coupling in Eq. (\ref{Dirac1}) does not affect the
density (\ref{c87.2}). The proportionality between temperature and
accelerations remains valid even if uniform accelerations are reached
asymptotically. The role of the anomalous magnetic moment in the energy
spectrum density will be discussed in a forthcoming publication.

\acknowledgments We thank Dr. Juan Rivero for helpful discussions.

\begin{figure}[tbp]
\centerline{\epsffile{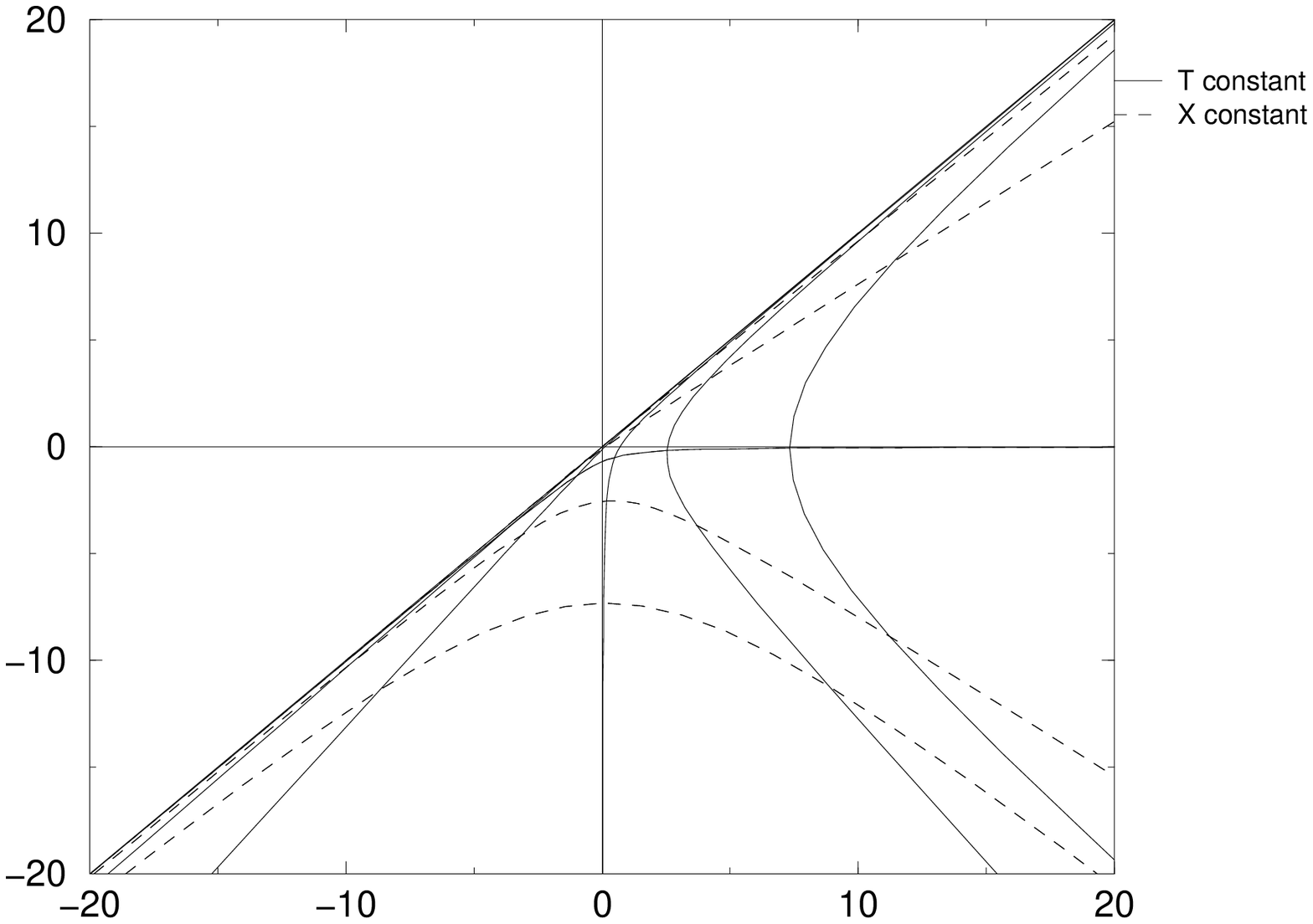}}
\caption{The accelerated coordinates with $\protect\omega=1$. The solid
lines correspond to the spacelike T=constant Cauchy surfaces. The dashed
lined correspond to the timelike curves X=constant}
\end{figure}

\end{document}